# New Policy Design for Food Accessibility to the People in Need

**Rahul Srinivas Sucharitha and Seokcheon Lee**
**School of Industrial Engineering, Purdue University,**
**West Lafayette, Indiana 47907, USA**

## Abstract

Food insecurity is a term used to measure hunger and food deprivation of a large population. As per the 2015 statistics provided by Feeding America - one of the largest domestic hunger-relief organizations in the United States, 42.2 million Americans live in food insecure households, including 29.1 million adults and 13.1 million children. This constitutes about 13.1% of households that are food insecure. Food Banks have been developed to improve food security for the needy. We have developed a novel food distribution policy using suitable welfare and poverty indices and functions. In this work, we propose an equitable and fair distribution of donated foods as per the demands and requirements of the people, thus ensuring minimum wastage of food (perishable and non-perishable) with focus towards nutrition. We present results and analysis based on the application of the proposed policy using the information of a local food bank as a case study. The results show that the new policy performs better than the current methods in terms of population being covered and reduction of food wastage obtaining suitable levels of nutrition.

**Keywords**
Humanitarian logistics, Policy design scheme, Optimization, Sustainability, Resource Planning

## 1. Introduction

Hunger though is a sensation that can be felt physically, it can be arduous to measure it. One such way to measure hunger and food deprivation of a large population is by food insecurity. Food insecurity is the state or condition of living without obtaining ample amounts of cost-effective, accessible and nourishing food. As per the 2015 statistics provided by Feeding America - one of the largest domestic hunger-relief organizations in the United States, 42.2 million Americans live in food insecure households, including 29.1 million adults and 13.1 million children. This constitutes about 13.1% of households that are food insecure [1]. To curb this issue, food insecure individuals and families receive assistance from the government and from philanthropic programs such as Feeding America. Feeding America provides food and assistance to these individuals by having a nation-wide network of around 200 food banks and around 60,000 food pantries and meal programs. Food banks obtain donated food and grocery products from food donors such as national food and grocery manufacturers, retailers, etc. The donated food is brought back to the food bank using either rented or owned trucks. These donated foods are processed in the food bank for quality purposes. The trucks are then used to distribute the donated foods to several food pantries and meal programs (termed as food agencies) based on their availability. The trucks return to the food bank once the donated foods are distributed as shown in Figure 1. The people receive the donated foods from these food pantries and meal programs.

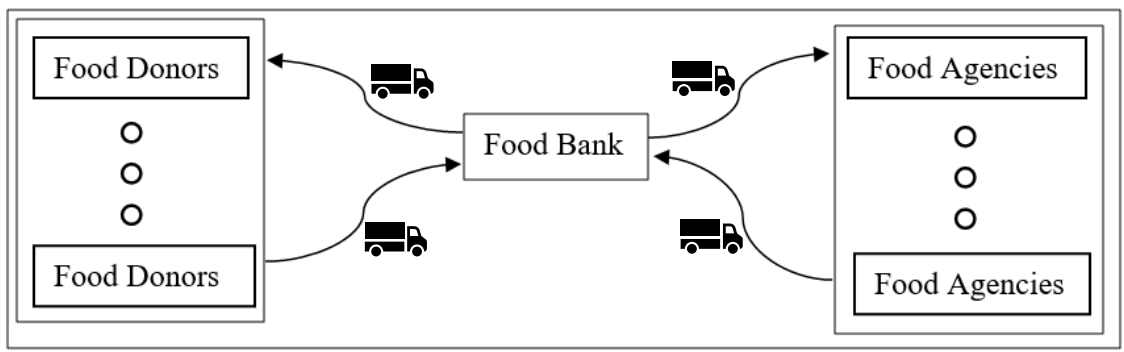

Figure 1: Current food distribution process of food banks

Food is usually received from food donors by following a fixed schedule that is convenient to the donors. Once the food is obtained from the food donors, the food is processed and inspected at the food bank before being made available for distribution to the food agencies. The food is distributed to the food agencies by trucks that are maintained by the food banks or is being rented based on the travel schedule. On conducting a field study with a local food bank, it was observed that the food agencies are first prioritized based on distance from the food bank (usually the food agencies nearest to the food bank are given priority). The amount of food provided is based on the storage capacity (in pounds) of each food agency. This leads to circumstances when, inadequate amount of nutritious food is provided to the people as the food is distributed based on poundage and not based on the poundage of the type of food (say protein, fats, etc.). Another issue that can be noticed in the distribution chain is the lack of correlation between the storage capacity of the food agencies and the total number of food insecure population in that area. This could lead to imbalance in feeding the intended population and increase in food wastage due to over-fulfilling a region with food that is more than required.

In this paper, we propose a decision policy scheme by implementing and providing an algorithm to the Local Food Bank authorities. This algorithm implements suitable poverty indices and social welfare functions to ensure people's needs are met during dynamic and uncertain situations and at the same time, ensuring distribution of nutritious food to everyone.

We use the preliminary information provided by a local food bank located in Lafayette, Indiana and compare our results with the current approach. Several food banks around the country lack sufficient information about the timing and availability of these food agencies. This leads to a dynamic environment and the usage of social welfare functions would aid in distributing the resources to the needy under uncertain circumstances.

The remainder of the paper is organized as follows. Section 2 provides a brief literature review. Section 3 presents the problem statement. The proposed approach is presented in section 4. Results and discussion is provided in section 5. Finally, section 6 concludes the paper and presents the potential future work.

## 2. Literature Review
### 2.1 Food Bank Literature
There have been some works that study the topic of food bank supply chain. [2][3] proposed mathematical models to facilitate the food bank's equitable and effective distribution of donated food. A linear programming model was formulated to maximize the effectiveness and a deviation constraint was formed to ensure perfect equity. Deterministic network-flow models were developed to minimize the amount of undistributed food. However, the model was proposed for a single food type.

From Previous research, it has been observed that there are several logistical issues that are being faced by non-profit organizations such as the food banks. Davis et al., [4] developed transportation schedules and allow food banks to collect food from the local food donors and deliver them to the food agencies. Food delivery points (FDPs) were developed by identifying satellite locations where agencies can receive food deliveries. A set covering model was developed to determine the food agency assignment to a FDP. In this model, vehicle capacity and food spoilage constraints were considered during the assignment. Using the optimal assignment of agencies to FDPs, schedules were constructed that considers the collection and distribution of donated food. However, the model does not consider the concept of a truck visiting multiple food delivery points food donors and excludes the study of various types of food and the distribution of different types of food to each region.

The donation of food and the demand of donated food is dynamic and uncertain in nature. It has been noticed that demand for food has been made deterministic. This has been taken as an assumption in several non-profit based distribution systems. Demand uncertainty and its estimation plays an important role in food bank operations as this can help us in understanding the number of people having food insecurity issues. For non-profit organizations, demand occurs in the form of supplies and people. Whereas, in for-profit organizations, Demand occurs in the form of products and services. Demand pattern also varies for both the organizations [5]. Supply of the donated goods can be done based on sufficient forecasting techniques. Supply for this kind of supply chain would be mainly dealing with ensuring sufficient inventory for the demand and studying the varying nature of the supply of the different types of donated goods (in-kind, etc.). Research related to demand and supply variability is limited in the non-profit based supply chain. To aid this issue of uncertainty, the U.S census bureau provides information of the people who are food insecure based on geography [6]. This information would benefit in the suitable forecasting of demand of donated food.

**2.2 Nutrition-based study**
Feeding America recently developed a set of guidelines for food banks titled "Foods to Encourage", with the main purpose of aiding the food bank officials with identifying nutritious food and making sure what kind of foods they should be aiming to distribute to the food agencies. [7] provides insights into the various nutrition-related policies and practices of food banks and food pantries and investigated on developments of online modules and assistance in food banks that aid in improving nutritional quality of food bank inventory.

The currently documented approach of donated food distribution among food banks and the recent research articles do not consider the concept of nutrition in their study and in their models. This needs to be taken into consideration as it has been observed through several studies that customers in food banks and food agencies prefer healthful foods and ranked less-healthful foods the lowest on their list of preference [7]. These results hence further reinforce the need for nutrition-focused food banks.

As per *MyPlate* guidelines [7] it is recommended that everyone divide his or her plate into approximately thirty percent grains, forty percent vegetables, ten percent fruits and twenty percent protein accompanied by small amounts of dairy. These portions of different types of food can be implemented in the social welfare functions to ensure distribution of equitable, effective and nutritious food to all the food agencies. We utilize these concepts in our study and observe the results.

**2.3 Inequality and Welfare functions**
Social welfare functions are used to assess various alternatives using certain inputs as preferences of the agents involved in the system [8]. There are many social welfare functions mentioned in previous literature. Among them, Atkinson family of inequality indices satisfies normalization, symmetry, population replication, principle of transfers, differentiability and scale independence [9]. Apart from this, the Atkinson social welfare function utilizes the concept of egalitarian equivalent income which is to be considered in a non-profit organization.

Most social welfare functions are defined considering the income inequality and the average income of the society. The social welfare functions are hence used to compare the income welfare in space and time. Similarly, combining the average and inequality of an income population, Atkinson [9], introduced a social welfare function which will be implemented in our research problem to measure the degree of equity and effectiveness.

## 3. Problem Statement

There is a growing need to ensure the distribution of necessary nutritious food to the needy. We examine the usage of Atkinson welfare function and the poverty risk index in the current distribution process of the food in the food bank supply chain.

The question that this research attempts to answer is: Is it possible to provide nutritious food in equitable amounts to the needy and at the same time ensure minimum food wastage? What if instead of reaching out to the nearest food agencies first and neglecting fuel costs, if we can first provide food to those agencies that are in dire need (based on poverty index calculations) would there be better equity of donated food among the agencies?

Considering that food banks process the food donated by the food donors, and with a good understanding of the inequality and welfare functions, the food banks can implement these techniques using currently available software at no cost to ensure equitable food distribution and optimum food distribution daily.

Given the nature of this problem, we examine the proposed approach of implementing suitable welfare functions and poverty indices to the current decision policy approach and compare the results with the current decision policy of donating food to the nearest food agencies from the food bank. As an example, to illustrate the research problem, we study a food bank in a region serving a group of food agencies with donated food. In the current approach, the food bank sends the food truck containing donated food to the food agencies, the first food agency obtaining the food is the one closest to the food bank. Also, the quantity of food donated to each food agency is based on their storage capacity. However, in view of the proposed decision policy approach, it is suggested to the food bank authorities to follow an algorithmic framework based on which food is first donated to those food agencies that are in dire need of food to provide to the needy people in their region, as compared to other food agencies and provide the right amount of food to each food agency based on the demand of the people, the nutrition requirement, and the storage capacity of the food agency.

## 4. Proposed Approach
The current approach to food distribution is like that of the nearest neighbor policy. Food trucks obtain the food from the food donors based on a fixed weekly schedule, and unload it in the food bank for processing and inspection. Once complete, the food trucks, either managed by the food banks or rented, are sent to the food agencies based on previous records of fixed schedules. The trucks visit the food banks based on their distance from the food bank. The agency closest to the food bank is visited first and this approach is continued until the food agencies demands are met or until the donated food is depleted for the day. By doing this, the people who need food are not provided with the donated food. Hence, in dynamic and uncertain conditions, implementing inequality and welfare functions for food distribution would prove to be effective to increase people's satisfaction and minimize food wastage.

### 4.1 Algorithm description
The algorithm is developed based on the application of suitable metrics that can measure the degree of equity and effectiveness. Such kind of metrics are obtained from economics in the form of inequality and welfare indices. For this research, we implement the adapted Atkinson social welfare function defined in [9] as follows:

$$R_f^x = \left[\frac{1}{n} \sum_{Fd \in F} (N_{Fd}^x)^{1-\epsilon_x}\right]^{\frac{1}{1-\epsilon_x}} \qquad (1)$$

Where $R_f^x$ is the resource welfare function, n is the number of food donors from a group of food donors F donating food to the food bank, $N_{Fd}^x$ is the residual of the resource x obtained from the food donor that is available for donation in the food bank, $\epsilon_x$ is the inequality aversion parameter for each type of resource x that will represent the penalty that each type of resource will take for each food agency based on the community food preferences. Since there are multiple food donors and multiple food types, the combined welfare function is defined as follows:

$$sR_f = \sum_{x=1}^{p} \beta^x R_f^x \qquad (2)$$

Where $\beta^x$ is the weight factor for the welfare of the type of resource and $\sum \beta^x = 1$. The weight is determined based on the nutrition requirement of each type of food that a person needs as mentioned in *MyPlate*. This is done to ensure nutrition focused food banking. For measuring the poverty level of each region surrounding the food agency, the most common head count ratio is implemented [8]. It is the ratio of the total number of people that are poor to the total number of population in that region. This information is obtained from the U.S Census Bureau data [6].

Based on the welfare functions and indices, the algorithm contains six main steps as follows:
  i. *Food donor arrangement*: The various food donors and their donated food are arranged in the order of decreasing perishability. For instance, F₁ = {f1, f2, f3} = {10,25,0} where, F₁ is the food donor donating f1, f2 and f3 food types to the food bank.
  ii. *Food agency arrangement*: The various food agencies that require the donated food to provide to the needy are arranged as per decreasing order of the poverty index P. the poverty index used is the head count ratio and the information needed to calculate this is obtained from the U.S Census Bureau data [6].
  iii. *Food donors-food agency evaluation*: From the list of ordered food agencies, the first one is chosen (say F$_a$ with certain food type demands) and the demand can be met with the single donor or a multiple set of food donors if it satisfies the requirement, $\sum F_i \geq F_a$. If not, then with the available food donors, the food is divided based on the available donations and the social welfare functions are implemented on all the possible combinations.
  iv. *Social welfare function implementation*: For each combination of the food donors for a food agency, social welfare function is implemented and the results are tabulated.
  v. *Food donors-food agency selection*: The combination with the greatest value of the combined welfare function is chosen and that combination is used for the food distribution.
  vi. *Execution and repeat*: Once a food agency's demand has been satisfied, the process is repeated for the next food agency in the list.

Given the size of this problem, the algorithm outlined above can be easily implemented in Microsoft Excel via Visual Basic for Applications. Since food banks are non-profit organizations, usage of this software would prove to be beneficial as it is available free of cost and with no additional purchase requirement. The policy designed in this paper is used to assist the authorities in the food banks in creating a more effective food distribution schedule. The output of this policy would provide the total number of people being fed and the amount of food overflow that would lead to food wastage.

## 5. Results and Discussion

The proposed approach is studied using a simple real-life case study and the performance is analyzed. We utilize the approach considering a time-period of one day because of the dynamic nature of the food distribution process of the food bank that varies on a day-to-day basis. Table 1 provides the information obtained for a single day from a local food bank. Since food banks lack sufficient personnel handling data, Table 1 details are the only data points that are provided by them. The data was recorded by following the current approach of distributing food to the nearest food agency and as per the capacity of the food agency. Although further information was unavailable, for the proposed approach, we take in certain assumptions and perform a preliminary simulation test. For the simulation of a single day time-period, we consider the given 10 food donors, 5 food agencies, and 1 food bank. Each food donor (arranged in the order of decreasing perishability of their donated food) will provide three types of food of varying amounts between 600 to 800 pounds each. The food agencies are arranged in the order of increasing distance from the food bank for the study of the current approach and later the food agencies are arranged in the order of increasing poverty index value (the information obtained from the U.S Census Bureau data of the given region) for the study of the proposed approach. Based on these arrangements, the first food agency is chosen. The food agency demands are randomly generated between 1,000 pounds and 2,000 pounds. The donated food is also split into three different food types. We consider a 50-km x 50-km region and the five food agencies are placed as per their current location in the region. The inequality aversion parameter is taken to be 1.5 and the weight factors considered for the three food types are taken as 1/3. The algorithm is implemented and the performance of both the approaches are described in Figure 3.

Table 1: Local food bank data

| Amount of Donated food obtained + inventory | 22280 pounds |
|---|---|
| Number of food agencies requesting for food | 5 |
| Total number of food donors | 10 |
| Total amount of food wasted | 8200 pounds |

The algorithm is compared to the current method of food distribution. In the current method, the donated food is distributed to food agencies based on distance and the capacity of the food agency. The performance is measured by the total amount of food overflow (or food wastage) in pounds and the number of people that have been served.

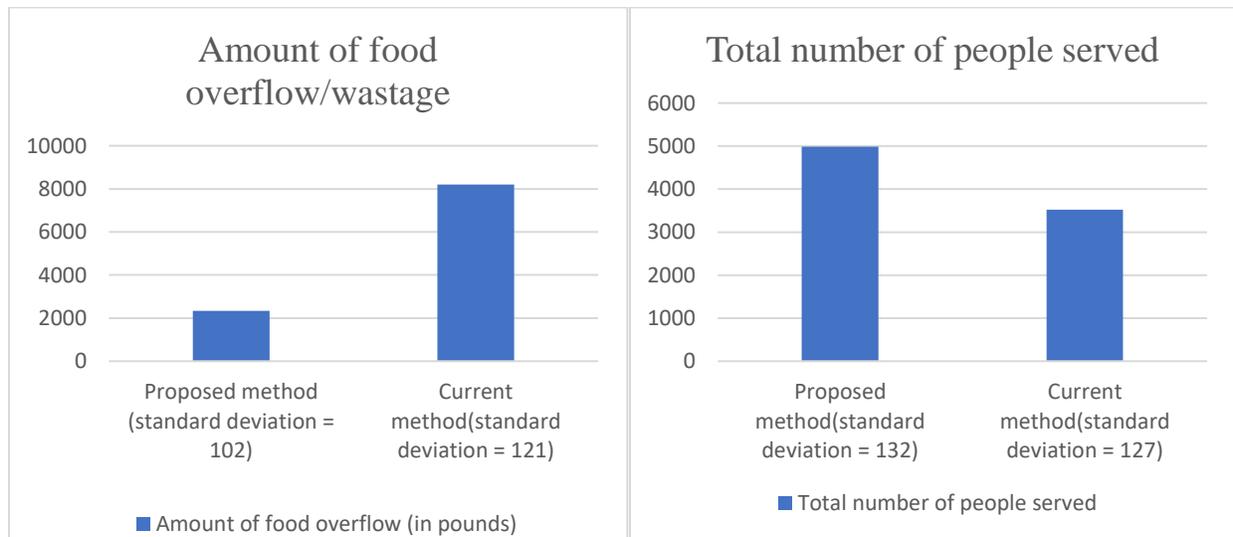

(a)　　　　　　　　　　　　　　　　　　　(b)

Figure 3: Comparison of performance of current method and proposed method

The simulation experiments are performed for multiple replications and the mean results are shown in Figure 3. Note that the proposed method performs better in preventing food overflow in food agencies as compared to the current method of donated food supply during a one-day period (Figure 3a) and serves more needy people than the current method as well (Figure 3b). On further studying the proposed method, tests were done to see if there would be changes in the performance measures by varying the inequality aversion parameter and comparing them with the current method. The results show that the proposed method outperforms the current method in various values of the inequality aversion parameter as can be seen in Figure 4.

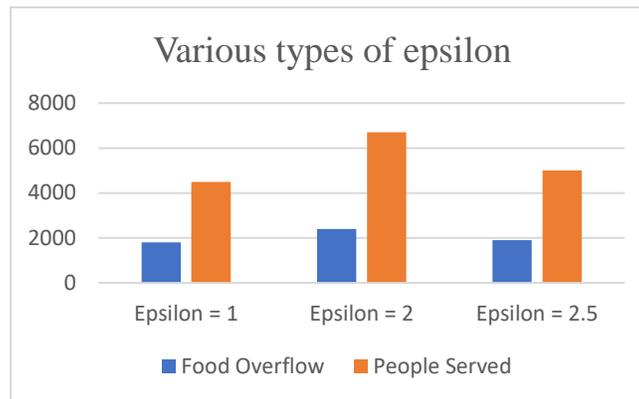

Figure 4: Comparison of performance of proposed method by varying inequality parameter

## 6. Conclusions and Future Work

This paper presents a new policy design for food banks based on social welfare and poverty indices. The algorithm proposed generated an equitable and fair distribution of donated foods as per the demands and requirements of the people in a simulated case study, thus ensuring minimum wastage of food (perishable and non-perishable) with focus towards nutrition. The results demonstrated fairer and better performance than the current approach of food distribution followed in food banks and these results will be showcased to the local food bank management to obtain their feedback and validation. This approach can be implemented on any food bank distribution structure and can include as many food types and nutritional constraints as required. It should also be noticed that due to lack of information, several assumptions were taken into consideration while performing the simulation study. Future work in this area will focus on the implementation of the algorithmic framework on varying time periods (weekly-based, monthly-based, etc.) and observe the results, followed by the development of a multi-objective mathematical model implementing the inequality and welfare functions and impact of individual and community nutrition preference of food and food wastage.

## References
1. "How Do Food Banks Work? | Feeding America® Food Banks." [Online]. Available: http://www.feedingamerica.org/about-us/how-we-work/food-bank-network/. [Accessed: 21-Mar-2017].
2. I. S. Orgut, J. Ivy, R. Uzsoy, and J. R. Wilson, "Modeling for the equitable and effective distribution of donated food under capacity constraints," IIE Trans., vol. 48, no. 3, pp. 252–266, Mar. 2016.
3. I. S. Orgut, J. Ivy, and R. Uzsoy, "Modeling for the Equitable and Effective Distribution of Food Donations under Stochastic Receiving Capacities."
4. Davis, L.B., Sengul, I., Ivy, J. S., Brock, L. G., & Miles, L. (2014). Scheduling food bank collections and deliveries to ensure food safety and improve access. Socio-Economic planning Sciences, 48(3), 175-188.
5. B. M. Beamon and B. Balcik, "Performance measurement in humanitarian relief chains," Int. J. Public Sect. Manag. Int. J. Phys. Distrib. Logist. Manag. Iss J. Manuf. Technol. Manag., vol. 21, no. 3, pp. 4–25, 2008.
6. "2010 Geographic Terms and Concepts - Census Tract – Geography - U.S. Census Bureau". Census.gov.Web. 13 Oct. 2016.
7. E. Campbell, K. Webb, M. Ross, P. Crawford, H. Hudson, and K. Hecht, "Nutrition-Focused Food Banking," 2015.
8. A. Villar, "Lecture Notes in Economics and Mathematical Systems 685 Lectures on Inequality, Poverty and Welfare."
9. Atkinson, A. B. (1970). On the measurement of inequality. Journal of Economy Theory, 2(3), 244-263.